%% file: main.tex
\documentclass[%
reprint,
superscriptaddress,
amsmath,amssymb,
aps,
table]{revtex4-2}
\usepackage{comment}
\usepackage{svg}
\usepackage{float}
\usepackage{mathtools}
\usepackage{siunitx}
\usepackage{graphicx}
\usepackage{dcolumn}
\usepackage{bm}
\usepackage{soul} 

\usepackage{graphicx}   
\usepackage{subcaption} 
\usepackage{hyperref}
\hypersetup{colorlinks = true,
            linkcolor = blue,
            urlcolor  = blue,
            citecolor = blue,
            anchorcolor = blue}

\usepackage{verbatim}
\usepackage[justification=raggedright,singlelinecheck=false]{caption}
\usepackage{float}
\usepackage{xspace}
\newcommand{\PI}{phase~I\xspace}
\newcommand{\PII}{phase~II\xspace}
\newcommand{\PIIa}{phase~II(a)\xspace}
\newcommand{\PIIb}{phase~II(b)\xspace}

\newcommand{\PIIcd}{phase~II(c) and II(d)\xspace}
\newcommand{\PIIab}{phase~II(a) and II(b)\xspace}
\newcommand{\PIIunpub}{phase~II(e)\xspace}

\newcommand{\PIIDPJointAgg}{$|\chi_{\text{rand}}|\geq2.90\times 10^{-15}$\xspace}

\newcommand{\PIIeJointAgg}{3.43$\times$$|g_{\gamma}^{\text{KSVZ}}|$\xspace}
\newcommand{\PIIeDPJointAgg}{$|\chi_{\text{rand}}|\geq 4.90\times10^{-15}$\xspace}
\newcommand{\PIIeDPSigma}{$17.1\sigma$\xspace}

\begin{document}

\preprint{APS/123-QED}
\title{Search for Dark Photons between 16.96--\SI{19.52}{\micro\eV} with the HAYSTAC Experiment}

\input{authors.tex}
\date{\today}

\begin{abstract}
We report dark photon results from HAYSTAC phase II using data from previously reported axion searches. Additionally, we present an analysis of an unpublished dataset covering a region between \SI{19.46}{}--\SI{19.52}{\micro\eV}. This region overlaps with a recently reported dark photon signal at \SI{19.5}{\micro\eV} with a kinetic coupling strength of $|\chi_{\text{rand}}| \simeq 6.5 \times 10^{-15}$ resulting from a reanalysis of previously published data from the TASEH collaboration. Given HAYSTAC's sensitivity, if such a signal were present, it would have appeared as a large \PIIeDPSigma excess above the noise. However, no such signal was observed. We thus exclude couplings \PIIeDPJointAgg at the 90\% confidence level over the newly reported region. In addition, using our previously reported axion data, we exclude couplings \PIIDPJointAgg between \SI{16.96}{}--\SI{19.46}{\micro\eV} at the 90\% confidence level.
\end{abstract}

\maketitle
The dark photon is a hypothetical massive vector boson, often considered a viable dark matter candidate alongside axions and axion-like particles~\cite{Arias2012}. Dark photons arise from the simplest form of a ``dark sector" model, which introduces an additional U(1) symmetry to the standard model. Aside from gravity, the only interaction between dark photons and standard model particles is kinetic mixing between dark photons and photons that would cause dark photons to oscillate to photons with mixing strength given by $\chi$~\cite{Fabbrichesi2020}.

Dark photon analyses can be carried out by reinterpreting existing axion exclusions from haloscope experiments~\cite{darkphoton_handbook,darkphoton_sumita}. Haloscopes consist of a tunable resonant cavity placed in a strong magnetic field, enabling axion-photon conversion where axions convert to photons via the inverse Primakoff effect~\cite{Sikivie:1983ip_halotheory,Sikivie:1985yu_halotheory}. These cavities are also sensitive to kinetic mixing between dark photons and photons, which, unlike axion-photon conversion, can take place with or without the magnetic field. 

The on resonance signal power, in natural units, for dark photons in a haloscope is given by
\begin{equation}
    P = \left(VC_{mnl}Q_\text{L}\frac{\beta}{1+\beta}\right)\left(\chi^2 m_{\chi}\rho_{\chi}\cos^2\theta\right) \label{eq:DP_power}.
\end{equation}
\noindent The first set of terms describes the properties of the haloscope, where $V$ is the cavity volume, $C_{mnl}$ is the form factor which quantifies the overlap between the cavity mode and the longitudinal axis, $Q_\text{L}$ is the cavity's loaded quality factor, and $\beta$ is the coupling strength of the readout antenna. The loaded quality factor is obtained from the unloaded value $Q_\text{0}$ through $Q_\text{L} = Q_\text{0} / (1 + \beta)$. The second set of terms describes the properties of the dark photon: $\chi$ is the strength of the kinetic mixing, $m_\chi$ is the dark photon mass, and $\rho_\chi = \SI{0.45}{\giga\electronvolt\per\centi\meter\cubed}$ is the local dark matter density~\cite{Read2014}. Finally, $\theta$ is the angle between the cavity's longitudinal axis and the dark photon's polarization. 

The TASEH collaboration released axion search results between 19.4687--\SI{19.8436}{\micro\eV} in 2022, excluding axion photon couplings $|g_{a\gamma\gamma}|\ge8.1\times10^{-14}$~\SI{}{G\eV^{-1}}~\cite{TASEH:2022}. Recently, collaborators of TASEH reanalyzed this data to search for dark photons and claimed a tentative dark photon signal at \SI{19.479727}{\micro\eV} (\SI{4.710178}{\GHz}) with a kinetic mixing strength of $|\chi_{\text{rand}}| \simeq 6.5 \times 10^{-15}$ assuming a randomly polarized dark photon field~\cite{TASEH_DP_2025}. This tentative signal was present during the original TASEH axion analysis but was excluded because it persisted in the absence of a magnetic field. This is a valid method to veto false signals in an axion search, but it does not apply to a dark photon search, where the conversion process is independent of the magnetic field strength. While the HAYSTAC experiment's previously published data did not include this region, HAYSTAC has collected data in the range 19.46--\SI{19.52}{\micro\eV} during a short run that was later aborted due to degraded cavity performance. Given the new results from TASEH, we have analyzed this data with careful treatment of the additional uncertainties in detector performance across this run. The results here include the analysis of this new region containing the TASEH candidate frequency, along with results for a dark photon search across all of HAYSTAC's phase II data.

\begin{figure*}[t]
    \centering
    \includegraphics[width=\linewidth]{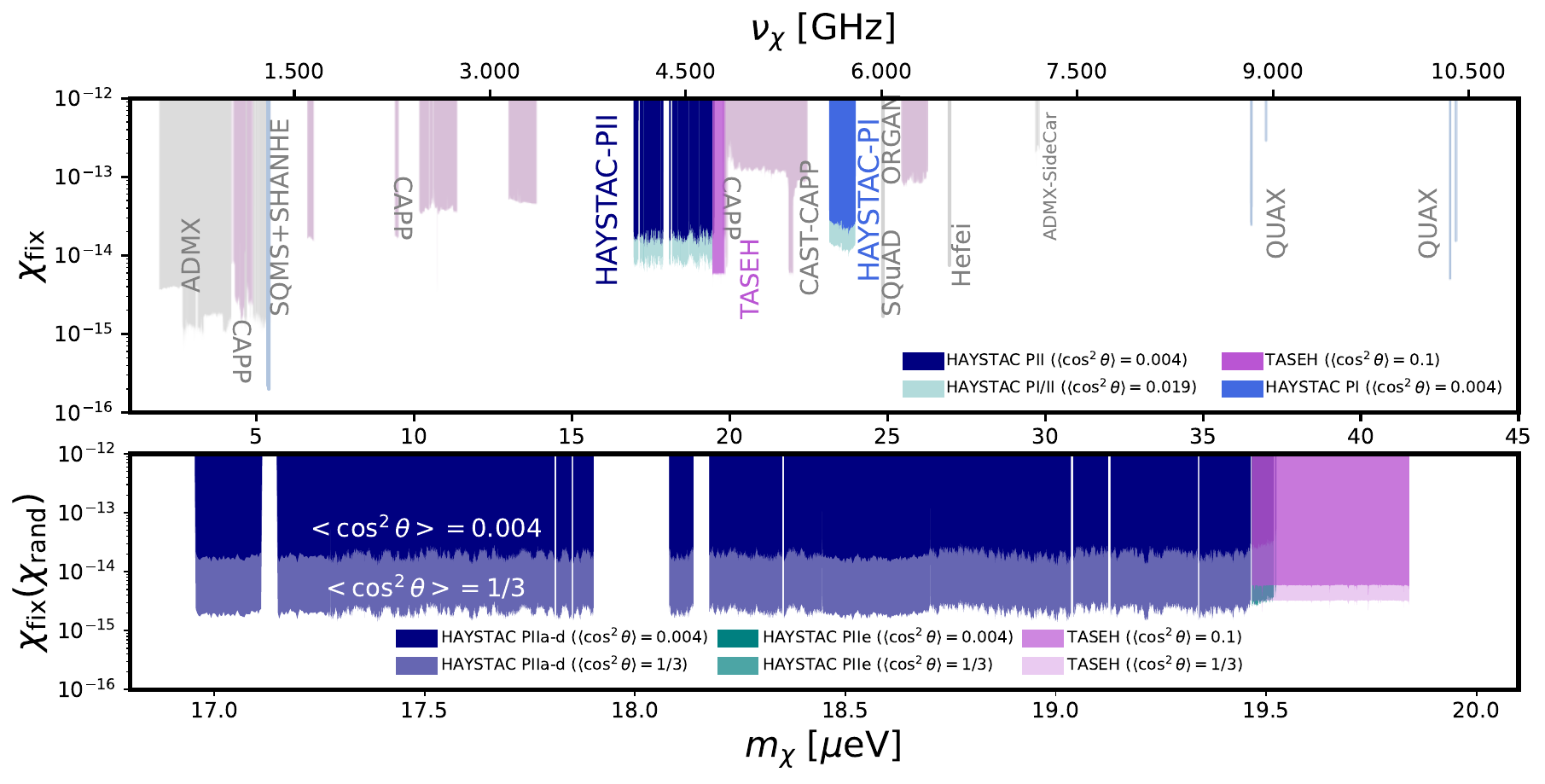}
    \caption{Top: Dark photon kinetic mixing for the fixed polarization scenario in the 1--\SI{45}{\mu eV} mass range, showing results from TASEH~\cite{TASEH:2022,TASEH_DP_2025}, ADMX~\cite{ADMX2024}, CAPP~\cite{CAPP8:2024, CAPPMAX:2024}, CAST-CAPP~\cite{CASTCAPP:2022},  ORGAN~\cite{ORGAN:2024}, QUAX~\cite{QUAX:2024}, SQuAD~\cite{SQuAD:2020ymh}, SQMS~\cite{SQMS:2022gtv}, Hefei~\cite{Hefei:2024slu}, SHANHE~\cite{SHANHE:2023kxz} (adapted from online repository~\cite{cajohare:github}). Also shown are HAYSTAC \PI and \PII, including exclusions converted by the method discussed in the text 
    which uses $\langle \cos^2\theta \rangle=0.004$ (blue), and what is currently shown in \cite{cajohare:github} (light green), which uses $\langle \cos^2\theta \rangle=0.019$ for a fair comparison with other results. Bottom: Exclusion on the dark photon kinetic mixing, comparing both the random (light fill) and fixed (solid fill) polarization scenarios,  highlighting HAYSTAC \PII (a)-(d) (blue), \PIIunpub (green) and TASEH (purple). The narrow vertical gaps mark the regions that are removed from the exclusion due to RFI contamination described in the text. } 
    \label{fig:exclusion}
\end{figure*}

HAYSTAC has run thus far in two unique phases. phase~I (January 2016--July 2017) scanned axion masses from \SI{23.15}{\micro\electronvolt}--\SI{24.0}{\micro\electronvolt} using a single Josephson Parametric Amplifier (JPA) with near quantum-limited noise \cite{brubaker2017analysis, zhong2018results}. In phase~II, the experiment was upgraded with a squeezed state receiver, leveraging squeezed vacuum states to achieve scan rate enhancements up to a factor of two relative to unsqueezed operation~\cite{malnou2019squeezed}. The upgraded experiment ran in subphases: \PIIa (September 2019--April 2020) which scanned \SI{16.96}{\micro\electronvolt}--\SI{17.28}{\micro\electronvolt}~\cite{backes2021quantum}, \PIIb (July 2021--November 2021) which scanned \SI{18.44}{\micro\electronvolt}--\SI{18.71}{\micro\electronvolt}~\cite{HAYSTAC_2023MJ}, and \PIIcd (September 2022--August 2024) which scanned \SI{17.28}{\micro\electronvolt}--\SI{18.44}{\micro\electronvolt} and \SI{18.71}{\micro\electronvolt}--\SI{19.46}{\micro\electronvolt}~\cite{Bai_HAYSTAC_2025}. Finally, a previously unpublished dataset, referred to as \PIIunpub, was collected in July 2022 and scanned \SI{19.46}{\micro\eV}--\SI{19.52}{\micro\eV}.

\begin{figure*}[t]
    \centering
    \begin{subfigure}{0.47\textwidth}
        \centering
        \includegraphics[width=\linewidth]{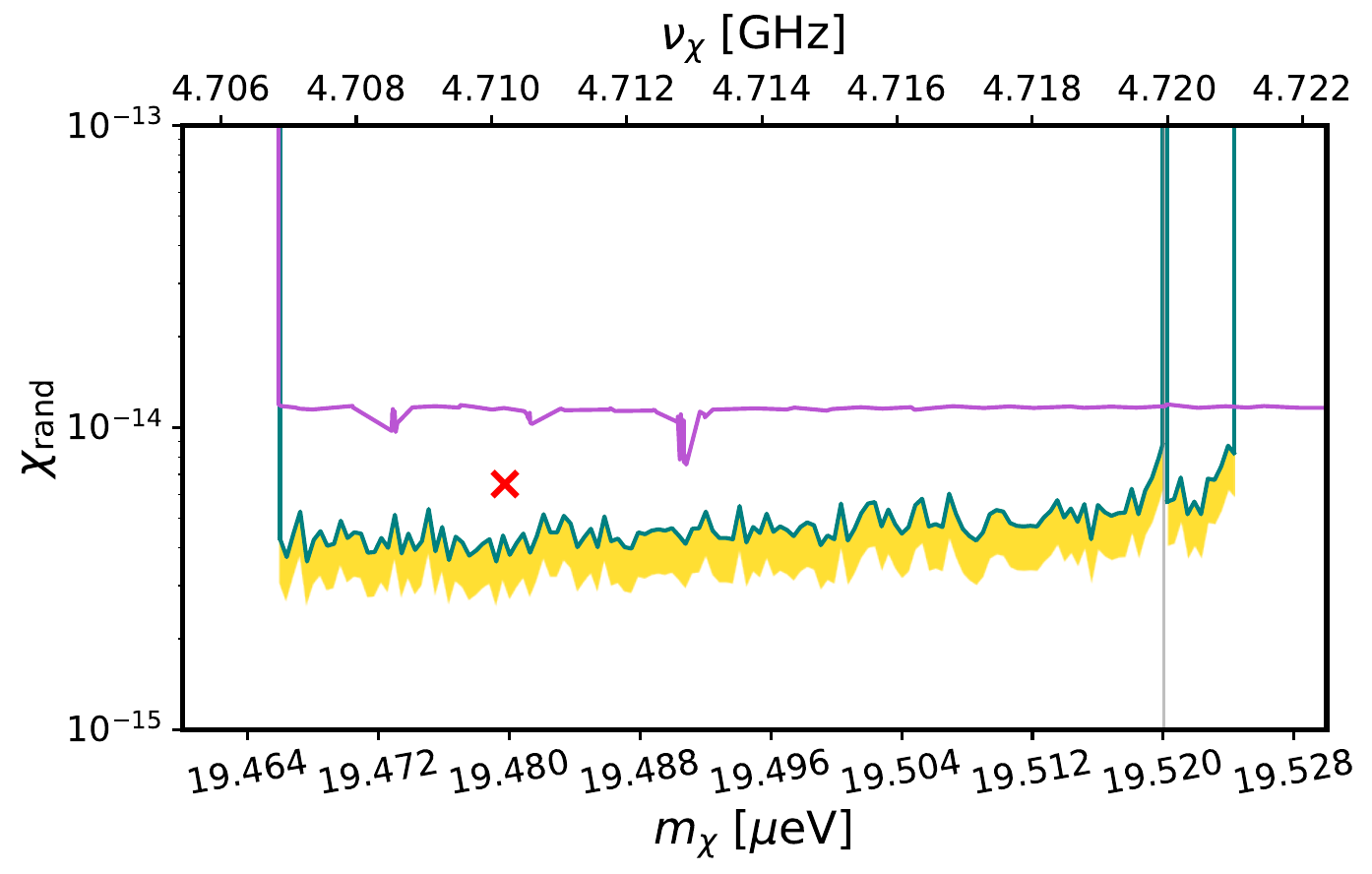}
    \end{subfigure}
    \hfill
    \begin{subfigure}{0.45\textwidth}
        \centering
        \includegraphics[width=\linewidth]{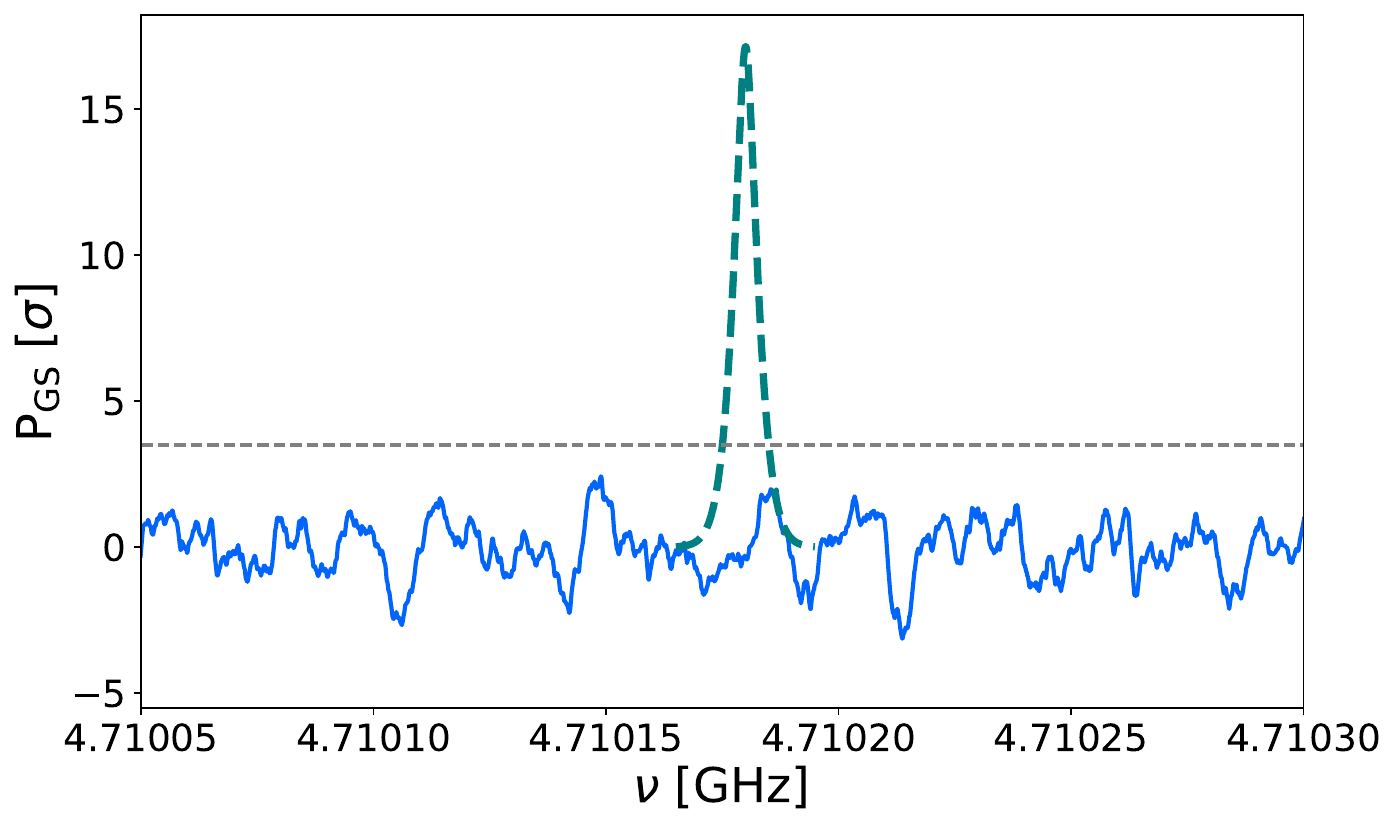}
    \end{subfigure}
    \caption{Left: Zoomed in exclusion on the dark photon coupling with random polarization showing results from HAYSTAC \PIIunpub in green and TASEH results in purple. The baseline exclusion, shown in green, assumes the conservatively degraded form factor, and the yellow shaded region highlights the degradation of the exclusion from the 50\% systematic uncertainty on the form factor. The red cross denotes the reported tentative dark photon signal from the TASEH reanalysis at \SI{19.479727}{\micro\eV} (\SI{4.710178}{\GHz}) with a kinetic mixing strength of $|\chi_\text{rand}| \simeq 6.5 \times 10^{-15}$~\cite{TASEH_DP_2025}. The vertical grey line marks the injected signal discussed in the text. Right: Normalized power observed in the grand spectrum P$_{\text{GS}}$  (blue solid line) at each observed frequency $\nu$, shown in a zoomed window around the reported dark photon signal. The horizontal grey dashed line corresponds to HAYSTAC's 3.468$\sigma$ rescan threshold. The green dashed peak shows the expected signal strength of \PIIeDPSigma, using the conservatively reduced form factor of 50\%, had the tentative signal been present during HAYSTAC's data taking.}
    \label{fig:exclusion_zoom}
\end{figure*}
The HAYSTAC experiment consists of a tunable copper-plated stainless steel cavity that is 25.4 cm long with an inner diameter of 10.2 cm. The tuning rod occupies 25\% of the cavity's total volume, leaving an unfilled volume of 1.545 L. The cavity is suspended from the mixing chamber, which is cooled to 60 mK, and placed inside an 8 T superconducting solenoidal magnet. HAYSTAC operates by observing the excess noise of the cavities $\text{TM}_{010}$ mode, which maximizes the overlap of the axion field with the magnetic field, giving an average C$_{010}$ of 0.44 over the frequency range of  \PII~(a)-(d) and 0.37 over \PIIunpub. An antenna coupled to the cavity detects the internal field, with coupling strength $\beta$ tuned by adjusting its insertion depth, which was maintained between 6--11 over \PII in order to optimize the scan rate with the squeezed state receiver~\cite{malnou2019squeezed}. The unloaded quality factor is measured to be on average $\sim$\SI{40000}{} over \PII~(a)-(d) and $\sim$\SI{39000}{} over \PIIunpub. The extracted signal is then amplified by the squeezed state receiver and digitized. Further experimental details are given in \cite{Kenany2017design}. 

Data taking follows the standard procedure described in~\cite{Bai_HAYSTAC_2025}, which uses an automated script that rotates the tuning rod to move the cavity's frequency by $\sim$80 kHz and than reoptimizes the squeezed state receiver at the new frequency. For each tuning step, the digitizer records data over integration time $\tau$ and computes the average power spectral density (PSD) by dividing the data into \SI{10}{ms} segments so as to achieve a frequency resolution of 100 Hz. For \PIIab and \PIIunpub, the integration time for each spectrum is 60 min, and for \PIIcd, it is reduced to 20 min to increase the scan rate. Each PSD is then filtered to remove structures wider than expected from the axion line shape with a predicted width of $\sim5$ kHz at the reported mass range~\cite{Turner_1990}. The spectra are then scaled by their signal-to-noise ratio, aligned in frequency, and summed to form a combined spectrum. The power from neighboring bins is then cross correlated with the axion line shape, and the resulting spectrum is corrected to account for correlations introduced by the filtering.  This gives a final grand spectrum, which is approximately Gaussian distributed with $\mu = 0$ and $\sigma = 1$ in the absence of a signal. 

This work builds on our previous dark photon results from \PI and \PIIa~\cite{darkphoton_sumita}, adding results from \PIIb, \PIIcd, and a previously unpublished dataset, \PIIunpub. \PIIunpub was collected between July 14th and July 25th, 2022, spanning the frequency range of 19.46--\SI{19.52}{\micro\eV} (4.707--4.721 GHz) and consisting of 145 total spectra. During \PIIunpub data taking, we observed a sharp drop in the quality factor when the cavity's $\text{TM}_{010}$ frequency was at $\sim$4.71 GHz, resulting in $Q_\text{0}\sim$\SI{20000}{}, roughly a factor of two lower than expected from previous measurements and simulations.  Due to the degraded performance, the run was aborted. Once at room temperature, the cavity was inspected and it was observed that the tuning rod had slipped and come in contact with the bottom end cap of the cavity. This was subsequently fixed, and upon cooling down the cavity $Q_0$ was observed to return to its nominal value of $\sim$\SI{40000}{}. Aside from the degraded Q, the system functioned normally, but due to the deviation from the nominal operating conditions and the uncertainty in how the rod position changing might impact the form factor, we chose not to include this dataset in our previous publication.  Because the tentative dark photon signal reported in \cite{TASEH_DP_2025} is well within the reach of HAYSAC's sensitivity, we have carefully reanalyzed the data with a study of potential uncertainty that could arise from the degraded cavity performance.

To quantify the effect of the rod's position changing during \PIIunpub, primarily on the form factor, we performed simulations in COMSOL~\cite{comsol_multiphysics_v63} and showed that the form factor C$_{010} = 0.41$ at the reported signal frequency \SI{4.710178}{\GHz} is highly insensitive to the vertical position of the rod in the cavity, varying by less than 1\% between the center position and the extreme upper or lower position.  Furthermore, we have confirmed that the broadening of a nearby mode crossing at $\sim$\SI{4.73}{GHz}, where there is a large dip in C$_{010}$, was too far away from the frequency of the reported signal to affect the C$_{010}$ there.  The reduction of the quality factor when the tuning rod slipped is a well known phenomenon due to the resulting high resistance and weak contact between the tuning rod and the bottom end cap. Out of prudence, for the results from \PIIunpub we assume a baseline form factor which is 2x lower than predicted in simulation, chosen to match the observed reduction in quality factor.

\PIIunpub data are analyzed using procedures described in~\cite{Bai_HAYSTAC_2025, HAYSTAC_2023MJ, brubaker2017analysis} and potential candidates are identified as power excesses exceeding 3.468$\sigma$, corresponding to a 10\% two-scan false negative rate for a target significance of 5.1$\sigma$ in the frequentist framework used in~\cite{brubaker2017analysis,brubaker2017analysis,zhong2018results}. Three candidates are identified at \SI{4.719720}{}, \SI{4.719816}{}, and \SI{4.719925}{\GHz}. The excess at \SI{4.719925}{\GHz} is part of the synthetic axion signal injection used for \PIIcd as described in~\cite{Bai_HAYSTAC_2025}, where we injected an RF signal through the cavity's weak port whose spectral shape matches the expected axion line shape~\cite{Zhu_2023}. This injected signal was correctly identified, and a data cut of 10~kHz was applied to remove data around the signal. Because the run was aborted, rescans were not performed on any of the candidates, and the data cut of the injected signal was not filled in as we did in \PIIcd.  The other two candidates were smaller excesses (P$_{\text{GS}}<4.5\sigma$) and did not correspond to any external sources. While also not rescanned, the Bayesian framework~\cite{palken2020improved} accounts for these power excesses by degrading the exclusion both at the observation frequency and of the overall aggregated exclusion.  This resulted in an exclusion on the axion-photon coupling of $|g_{\gamma}| >$ \PIIeJointAgg at the 90\% confidence level between 19.46--\SI{19.52}{\micro\eV} in \PIIunpub.  

Because \PIIcd were previously analyzed to search for axions, careful treatment of certain candidates is required to ensure no axion specific vetos, such as the magnetic field veto, have been used. This applies specifically to 12 excesses later identified as radio frequency interference (RFI), summarized in Table~II of~\cite{Bai_HAYSTAC_2025}, 5 of which persisted after ramping down the magnetic field. However, all excesses were found to correlate with signals detected by an ambient RF receiver, a simple room-temperature antenna located in the same room as the main detector, ruling them out as dark photon candidates. Persistent RFIs were removed from the data by applying a 10~kHz cut around each affected frequency.

 Given no signal was observed during \PII, the limits on axion photon coupling $g_{a\gamma\gamma}$ can be recast to limits on dark photon kinetic mixing $\chi$, using the prescription described in~\cite{darkphoton_handbook}
\begin{equation}
    \chi = g_{a\gamma\gamma}\frac{B_0}{m_{\chi}\sqrt{<\cos^2\theta>}},
    \label{eq:DP_recast}
\end{equation}
\noindent where $B_0$=\SI{1.56e-15}{GeV^2} is the magnetic field in the experiment in natural units. Since the polarization is unknown, two scenarios are considered in order to convert axion limits into dark photon limits~\cite{darkphoton_handbook}. In the first scenario, the polarization of the dark photon field is random and changes every coherence time $\tau_{\mathrm{coh}}$, which is $\sim \SI{200}{\micro\second}$ at the masses we are probing. In this case, as long as the $\tau\gg \tau_{\mathrm{coh}}$, we expect $\cos^2\theta = 1/3$. Using this, we can convert the axion exclusion from \PII~(a)-(d) to a dark photon limit with an aggregated exclusion of \PIIDPJointAgg at the 90\% confidence level between 18.71 -- \SI{19.46}{\micro\eV}. The same is done for \PIIunpub, which we treat separately due to a conservative assumption about the form factor degradation described above, giving an aggregated exclusion of \PIIeDPJointAgg at the 90\% confidence level between 19.46--\SI{19.52}{\micro\eV}. 

In the alternative scenario where the dark photon has a single fixed polarization, the value of $\cos\theta$ not only depends on the polarization angle itself, but also on the experiment’s location on Earth and its orientation relative to Earth’s rotation. Furthermore, $\cos\theta$ varies in time as the relative orientation between Earth and the dark photon field changes. Thus, converting an axion limit into a dark photon limit requires computing an effective value of $\cos\theta$ that preserves the statistical inference made in the original axion exclusion. HAYSTAC is located at a latitude of \ang{41.32} with the cavity's longitudinal axis pointing along the radius of the Earth. Prior to \PII, HAYSTAC performed its analysis using the frequentist framework described in~\cite{brubaker2017analysis}, where the exclusion limit is defined as the coupling strength for which, given a signal 5.1$\sigma$ above the noise, the probability that the observed power (P$_{\text{obs}}$) falls below the rescan threshold is 10\% after two scans. In this case, the effective $\langle \cos^2\theta \rangle$ for an instantaneous measurement is determined by solving for the normalized signal power $P_{\text{sig}}$ via simulation using a conditional probability relation, $\text{Prob}\!\left(P_{\mathrm{obs}} > 3.468\sigma \,\middle|\, {P_\mathrm{sig}} = 5.1\sigma / \langle \cos^2\theta \rangle\right) = \sqrt{0.9}$, which yields $\langle \cos^2\theta \rangle = 5.1\sigma / P_{\text{sig}} \approx 0.004$.  Note that this is different from the assumption made in~\cite{cajohare:github,darkphoton_handbook}, which estimated the dark photon signal strength that would exceed the mean noise 90$\%$ of the time, giving  $\langle \cos^2\theta \rangle = 0.019$.  As stated in \cite{darkphoton_handbook}, this is a practical assumption for converting a generic experiment, but it does not correspond to the actual analysis used by HAYSTAC. 

It should be noted that starting with \PII, HAYSTAC adopted a Bayesian framework, in which the exclusion is instead defined as the coupling strength for which the probability of an axion existing falls below 10\% given the observed power. In this case, because the exclusion is no longer a binary function of the observed power, the effective $\cos\theta$ value is not a constant across frequencies but varies in each frequency bin depending on the observed power. However, for the purpose of this work, we adopt the former value, $\langle \cos^2\theta \rangle \sim 0.004$, as a straightforward yet conservative choice. Under this assumption, the exclusions from \PII~(a)-(e) are converted to dark photon limits giving aggregated exclusion of $|\chi_{\text{fix}}|\geq2.64\times10^{-14}$ for  \PII~(a)-(d) and $|\chi_{\text{fix}}|\geq4.47\times10^{-14}$ for  \PII~(e) at the 90\% confidence level. The resulting HAYSTAC dark photon limits, along with those from other experiments under both random and fixed polarization scenarios, are shown in Fig.~\ref{fig:exclusion}.

In addition, given the overlap between \PIIunpub and the tentative signal reported from TASEH's reanalysis, we carefully analyzed our data to understand the expected response to such a signal. At the candidate frequency, HAYSTAC collected 27 spectra with varying sensitivity, corresponding to a total integration time of 27~hours. Assuming a randomly polarized dark photon field and adopting a conservative value for the form factor, the resulting 5.1$\sigma$ sensitivity corresponds to $|\chi_{\text{rand}}| = 3.54\times10^{-15}$ at the signal frequency. Under these conditions, the reported signal would be expected to appear as a $ (6.48/3.54)^2 \times 5.1\sigma \simeq 17.1\sigma$ excess above the noise. As shown in Fig.~\ref{fig:exclusion_zoom} (right), no such excess is observed. It should be noted that to match the analysis used in the tentative claim, the analysis here assumes the random polarization scenario, which is not impacted by variations in $\cos\theta$. If instead the dark photon field polarization is fixed, the signal in HAYSTAC could have been suppressed.  While we do not present a detailed analysis of this scenario, we will note that after accounting for deadtime, the 27 contributing spectra correspond to a total elapsed observation time of $\sim$40 hours with varying sensitivity in each scan. This is longer than one sidereal day, which reduces the effect from potential misalignment between the cavity axis and the dark photon field's polarization angle as detailed in \cite{darkphoton_handbook}.

In summary, we report dark photon results using HAYSTAC's phase~II axion search data, including a new search region from 19.46--\SI{19.52}{\micro\electronvolt}. This region overlaps with a recently claimed signal in a reanalysis of TASEH's data near \SI{19.5}{\micro\electronvolt}. Given HAYSTAC’s sensitivity, assuming a randomly polarized dark photon field, such a signal would have appeared well above the noise, but none was observed. We therefore exclude couplings at \PIIeDPJointAgg with 90\% confidence in this region. In addition, for the rest of \PII, we exclude couplings \PIIDPJointAgg between \SI{16.96}{}--\SI{19.46}{\micro\eV} at the 90\% confidence level.

\hfill \break
\textit{Acknowledgments}---HAYSTAC is supported by the National Science Foundation under grants PHY-2514170, PHY-1306729, the Heising-Simons Foundation under grants 2014-0904, 2016-044, and the David and Lucile Packard Foundation grant number 2022-74683. We thank Kyle Thatcher and Calvin Schwadron for their work on the design and fabrication of the squeezed state receiver mechanical components, Felix Vietmeyer for his work on the room temperature electronics, and Steven Burrows for his graphical design work. We thank Vincent Bernardo and the J.~W.~Gibbs Professional Shop as well as Craig Miller and Dave Johnson for their assistance with fabricating the system’s mechanical components. We also wish to thank the Cory Hall Machine shop at UC Berkeley and the efforts of Sergio Velazquez for the microwave cavity used in this run. We thank Dr.~Matthias Buehler of low-T Solutions for cryogenics advice. Finally, we thank the Yale Wright laboratory for housing the experiment and providing computing and facilities support.    

\appendix

\bibliography{apssamp,refs_haloscopes}
\end{document}

%% file: authors.tex
\newcommand{\Location}{\affiliation{Location}}
\newcommand{\Yale}{\affiliation{Department of Physics, Yale University, New Haven, Connecticut 06520, USA}}
\newcommand{\YaleApplied}{\affiliation{Department of Applied Physics, Yale University, New Haven, Connecticut 06520, USA}}
\newcommand{\WrightLab}{\affiliation{Wright Laboratory, Department of Physics, Yale University, New Haven, Connecticut 06520, USA}}
\newcommand{\Cal}{\affiliation{Department of Nuclear Engineering, University of California Berkeley, California 94720, USA}}
\newcommand{\JILA}{\affiliation{JILA, National Institute of Standards and Technology and the University of Colorado, Boulder, Colorado 80309, USA}}
\newcommand{\Colorado}{\affiliation{Department of Physics, University of Colorado, Boulder, Colorado 80309, USA}}
\newcommand{\Hopkins}{\affiliation{Department of Physics and Astronomy, The Johns Hopkins University, Baltimore, MD, 21218}}
\newcommand{\NIST}{\affiliation{National Institute of Standards and Technology, Boulder, Colorado 80305, USA}}

\author{Xiran~Bai}\Yale\WrightLab
\author{A.~Droster}\Cal
\author{J.~Echevers}\Cal
\author{Maryam H.~Esmat}\Hopkins
\author{Sumita~Ghosh}\altaffiliation{now at Lawrence Livermore Nation Laboratory}\YaleApplied\WrightLab
\author{Eleanor~Graham}\Yale\WrightLab
\author{H.~Jackson}\Cal
\author{S. Jois}\Cal
\author{M.J.~Jewell}\Yale\WrightLab
\author{Claire~Laffan}\Yale\WrightLab
\author{A.F.~Leder}\altaffiliation{now at Los Alamos National Laboratory}\Cal
\author{K.W.~Lehnert}\Yale\WrightLab
\author{S.M.~Lewis}\altaffiliation{now at Wellesley College}\Cal
\author{R.H.~Maruyama}\Yale\WrightLab
\author{N. M. Rapidis}\altaffiliation{now at Stanford University}\Cal
\author{E.P.~Ruddy}\Colorado\JILA
\author{M.~Silva-Feaver}\Yale\WrightLab
\author{M.~Simanovskaia}\altaffiliation{now at Stanford University}\Cal
\author{Sukhman~Singh}\Yale\WrightLab
\author{D.H.~Speller}\Hopkins
\author{K.~van~Bibber}\Cal
\author{Y.~Wang}\Hopkins
\author{Sabrina Zacarias}\altaffiliation{now at UC Santa Barbra}\Yale\WrightLab
\author{Yuqi~Zhu}\altaffiliation{now at Stanford University}\Yale\WrightLab

\collaboration{HAYSTAC Collaboration}

%% file: apssamp.bib
@misc{comsol_multiphysics_v63,
  author = {{COMSOL Multiphysics}},
  title = {{COMSOL Multiphysics v. 6.3}},
  howpublished = {www.comsol.com},
  note = {{COMSOL} AB, Stockholm, Sweden}
}

@article{darkphoton_sumita,
    author = "Ghosh, Sumita and Ruddy, E. P. and Jewell, Michael J. and Leder, Alexander F. and Maruyama, Reina H.",
    title = "{Searching for dark photons with existing haloscope data}",
    eprint = "2104.09334",
    archivePrefix = "arXiv",
    primaryClass = "hep-ph",
    doi = "10.1103/PhysRevD.104.092016",
    journal = "Phys. Rev. D",
    volume = "104",
    number = "9",
    pages = "092016",
    year = "2021"
}

@article{darkphoton_handbook,
    author = "Caputo, Andrea and Millar, Alexander J. and O'Hare, Ciaran A. J. and Vitagliano, Edoardo",
    title = "{Dark photon limits: A handbook}",
    eprint = "2105.04565",
    archivePrefix = "arXiv",
    primaryClass = "hep-ph",
    reportNumber = "NORDITA-2021-036",
    doi = "10.1103/PhysRevD.104.095029",
    journal = "Phys. Rev. D",
    volume = "104",
    number = "9",
    pages = "095029",
    year = "2021"
}

@article{Turner_1990,
    author = "Turner, Michael S.",
    title = "{Periodic signatures for the detection of cosmic axions}",
    reportNumber = "FERMILAB-PUB-90-089-A",
    doi = "10.1103/PhysRevD.42.3572",
    journal = "Phys. Rev. D",
    volume = "42",
    pages = "3572--3575",
    year = "1990"
}

@article{Zhu_2023,
    author = "Zhu, Yuqi and Jewell, M. J. and Laffan, Claire and Bai, Xiran and Ghosh, Sumita and Graham, Eleanor and Cahn, S. B. and Maruyama, Reina H. and Lamoreaux, S. K.",
    title = "{An improved synthetic signal injection routine for the Haloscope At Yale Sensitive To Axion Cold dark matter (HAYSTAC)}",
    eprint = "2212.00732",
    archivePrefix = "arXiv",
    primaryClass = "physics.ins-det",
    reportNumber = "FERMILAB-PUB-22-932-SQMS-V",
    doi = "10.1063/5.0137870",
    journal = "Rev. Sci. Instrum.",
    volume = "94",
    number = "5",
    pages = "054712",
    year = "2023"
}

@article{HAYSTAC_2023MJ,
    author = "Jewell, M. J. and others",
    collaboration = "HAYSTAC",
    title = "{New results from HAYSTAC\textquoteright{}s phase II operation with a squeezed state receiver}",
    eprint = "2301.09721",
    archivePrefix = "arXiv",
    primaryClass = "hep-ex",
    doi = "10.1103/PhysRevD.107.072007",
    journal = "Phys. Rev. D",
    volume = "107",
    number = "7",
    pages = "072007",
    year = "2023",
}

@article{Sikivie:1983ip_halotheory,
    author = "Sikivie, P.",
    editor = "Srednicki, M. A.",
    title = "{Experimental Tests of the Invisible Axion}",
    reportNumber = "PRINT-83-0597 (FLORIDA), UF-TP-83-13",
    doi = "10.1103/PhysRevLett.51.1415",
    journal = "Phys. Rev. Lett.",
    volume = "51",
    pages = "1415--1417",
    year = "1983",
    note = "[Erratum: Phys.Rev.Lett. 52, 695 (1984)]"
}

@article{Sikivie:1985yu_halotheory,
    author = "Sikivie, Pierre",
    title = "{Detection Rates for 'Invisible' Axion Searches}",
    reportNumber = "UFTP-85-5-REV, UFTP-85-5",
    doi = "10.1103/PhysRevD.36.974",
    journal = "Phys. Rev. D",
    volume = "32",
    pages = "2988",
    year = "1985",
    note = "[Erratum: Phys.Rev.D 36, 974 (1987)]"
}

@article{backes2021quantum,
    author = "Backes, K. M. and others",
    collaboration = "HAYSTAC",
    title = "{A quantum-enhanced search for dark matter axions}",
    eprint = "2008.01853",
    archivePrefix = "arXiv",
    primaryClass = "quant-ph",
    doi = "10.1038/s41586-021-03226-7",
    journal = "Nature",
    volume = "590",
    number = "7845",
    pages = "238--242",
    year = "2021"
}

@article{malnou2019squeezed,
    author = "Malnou, M. and Palken, D. A. and Brubaker, B. M. and Vale, Leila R. and Hilton, Gene C. and Lehnert, K. W.",
    title = "{Squeezed vacuum used to accelerate the search for a weak classical signal}",
    eprint = "1809.06470",
    archivePrefix = "arXiv",
    primaryClass = "quant-ph",
    doi = "10.1103/PhysRevX.9.021023",
    journal = "Phys. Rev. X",
    volume = "9",
    number = "2",
    pages = "021023",
    year = "2019",
    note = "[Erratum: Phys.Rev.X 10, 039902 (2020)]"
}

@article{zhong2018results,
    author = "Zhong, L. and others",
    collaboration = "HAYSTAC",
    title = "{Results from phase 1 of the HAYSTAC microwave cavity axion experiment}",
    eprint = "1803.03690",
    archivePrefix = "arXiv",
    primaryClass = "hep-ex",
    doi = "10.1103/PhysRevD.97.092001",
    journal = "Phys. Rev. D",
    volume = "97",
    number = "9",
    pages = "092001",
    year = "2018"
}

@article{brubaker2017analysis,
    author = "Brubaker, B. M. and Zhong, L. and Lamoreaux, S. K. and Lehnert, K. W. and van Bibber, K. A.",
    title = "{HAYSTAC axion search analysis procedure}",
    eprint = "1706.08388",
    archivePrefix = "arXiv",
    primaryClass = "astro-ph.IM",
    doi = "10.1103/PhysRevD.96.123008",
    journal = "Phys. Rev. D",
    volume = "96",
    number = "12",
    pages = "123008",
    year = "2017"
}

@article{palken2020improved,
    author = "Palken, D. A. and others",
    title = "{Improved analysis framework for axion dark matter searches}",
    eprint = "2003.08510",
    archivePrefix = "arXiv",
    primaryClass = "astro-ph.IM",
    doi = "10.1103/PhysRevD.101.123011",
    journal = "Phys. Rev. D",
    volume = "101",
    number = "12",
    pages = "123011",
    year = "2020"
}

@article{Read2014,
    author = "Read, J. I.",
    title = "{The Local Dark Matter Density}",
    eprint = "1404.1938",
    archivePrefix = "arXiv",
    primaryClass = "astro-ph.GA",
    reportNumber = "JPHYSG-100038.R1",
    doi = "10.1088/0954-3899/41/6/063101",
    journal = "J. Phys. G",
    volume = "41",
    pages = "063101",
    year = "2014"
}

@article{Kenany2017design,
    author = "Al Kenany, S. and others",
    title = "{Design and operational experience of a microwave cavity axion detector for the 20\textendash{}100$\mu$eV range}",
    eprint = "1611.07123",
    archivePrefix = "arXiv",
    primaryClass = "physics.ins-det",
    doi = "10.1016/j.nima.2017.02.012",
    journal = "Nucl. Instrum. Meth. A",
    volume = "854",
    pages = "11--24",
    year = "2017"
}

@misc{cajohare:github,
  author    = {Ciaran O'Hare},
  title     = {cajohare/AxionLimits: AxionLimits (v1.0)},
  version   = {1.0},
  publisher = {Zenodo},
  year      = 2020,
  doi       = {10.5281/zenodo.3932430},
  url       = {https://doi.org/10.5281/zenodo.3932430}
}


%% file: refs_haloscopes.bib
@article{SQuAD:2020ymh,
    author = "Dixit, Akash V. and Chakram, Srivatsan and He, Kevin and Agrawal, Ankur and Naik, Ravi K. and Schuster, David I. and Chou, Aaron",
    title = "{Searching for Dark Matter with a Superconducting Qubit}",
    eprint = "2008.12231",
    archivePrefix = "arXiv",
    primaryClass = "hep-ex",
    reportNumber = "FERMILAB-PUB-20-424-E",
    doi = "10.1103/PhysRevLett.126.141302",
    journal = "Phys. Rev. Lett.",
    volume = "126",
    number = "14",
    pages = "141302",
    year = "2021"
}

@article{SHANHE:2023kxz,
    author = "Tang, Zhenxing and others",
    collaboration = "SHANHE",
    title = "{First Scan Search for Dark Photon Dark Matter with a Tunable Superconducting Radio-Frequency Cavity}",
    eprint = "2305.09711",
    archivePrefix = "arXiv",
    primaryClass = "hep-ex",
    doi = "10.1103/PhysRevLett.133.021005",
    journal = "Phys. Rev. Lett.",
    volume = "133",
    number = "2",
    pages = "021005",
    year = "2024"
}

@article{Hefei:2024slu,
    author = "Kang, Runqi and Jiao, Man and Tong, Yu and Liu, Yang and Zhong, Youpeng and Cai, Yi-Fu and Zhou, Jingwei and Rong, Xing and Du, Jiangfeng",
    title = "{Near-quantum-limited haloscope search for dark-photon dark matter enhanced by a high-Q superconducting cavity}",
    eprint = "2404.12731",
    archivePrefix = "arXiv",
    primaryClass = "hep-ex",
    doi = "10.1103/PhysRevD.109.095037",
    journal = "Phys. Rev. D",
    volume = "109",
    number = "9",
    pages = "095037",
    year = "2024"
}

@article{SQMS:2022gtv,
    author = "Cervantes, Raphael and others",
    title = "{Deepest sensitivity to wavelike dark photon dark matter with superconducting radio frequency cavities}",
    eprint = "2208.03183",
    archivePrefix = "arXiv",
    primaryClass = "hep-ex",
    reportNumber = "FERMILAB-PUB-22-584-SQMS",
    doi = "10.1103/PhysRevD.110.043022",
    journal = "Phys. Rev. D",
    volume = "110",
    number = "4",
    pages = "043022",
    year = "2024"
}

@article{Bai_HAYSTAC_2025,
    author = "Bai, Xiran and others",
    collaboration = "HAYSTAC",
    title = "{Dark Matter Axion Search with HAYSTAC Phase II}",
    eprint = "2409.08998",
    archivePrefix = "arXiv",
    primaryClass = "hep-ex",
    doi = "10.1103/PhysRevLett.134.151006",
    journal = "Phys. Rev. Lett.",
    volume = "134",
    number = "15",
    pages = "151006",
    year = "2025"
}

@article{ADMX2024,
    author = "Bartram, C. and others",
    collaboration = "ADMX",
    title = "{Axion Dark Matter eXperiment around 3.3 $\mu$eV with Dine-Fischler-Srednicki-Zhitnitsky Discovery Ability}",
    eprint = "2408.15227",
    archivePrefix = "arXiv",
    primaryClass = "hep-ex",
    reportNumber = "FERMILAB-PUB-24-0602-V",
    year = "2024",
    journal = "Submitted to Phys. Rev. Lett."
}

@article{CAPP8:2024,
    author = "Kim, Younggeun and others",
    title = "{Experimental Search for Invisible Dark Matter Axions around 22\,\,\ensuremath{\mu}eV}",
    eprint = "2312.11003",
    archivePrefix = "arXiv",
    primaryClass = "hep-ex",
    doi = "10.1103/PhysRevLett.133.051802",
    journal = "Phys. Rev. Lett.",
    volume = "133",
    number = "5",
    pages = "051802",
    year = "2024"
}

@article{CAPPMAX:2024,
    author = "Ahn, Saebyeok and others",
    collaboration = "CAPP",
    title = "{Extensive Search for Axion Dark Matter over 1~GHz with CAPP{\textquoteright}S Main Axion Experiment}",
    eprint = "2402.12892",
    archivePrefix = "arXiv",
    primaryClass = "hep-ex",
    doi = "10.1103/PhysRevX.14.031023",
    journal = "Phys. Rev. X",
    volume = "14",
    number = "3",
    pages = "031023",
    year = "2024"
}

@article{CASTCAPP:2022,
        author = "Adair, C. M. and others",
    title = "{Search for Dark Matter Axions with CAST-CAPP}",
    eprint = "2211.02902",
    archivePrefix = "arXiv",
    primaryClass = "hep-ex",
    doi = "10.1038/s41467-022-33913-6",
    journal = "Nature Commun.",
    volume = "13",
    number = "1",
    pages = "6180",
    year = "2022"
}

@article{TASEH:2022,
      author = "Chang, Hsin and others",
    collaboration = "TASEH",
    title = "{First Results from the Taiwan Axion Search Experiment with a Haloscope at 19.6\,\,\ensuremath{\mu}eV}",
    eprint = "2205.05574",
    archivePrefix = "arXiv",
    primaryClass = "hep-ex",
    doi = "10.1103/PhysRevLett.129.111802",
    journal = "Phys. Rev. Lett.",
    volume = "129",
    number = "11",
    pages = "111802",
    year = "2022"
}

@article{QUAX:2024,
    author = "Rettaroli, A. and others",
    collaboration = "QUAX",
    title = "{Search for axion dark matter with the QUAX\textendash{}LNF tunable haloscope}",
    eprint = "2402.19063",
    archivePrefix = "arXiv",
    primaryClass = "physics.ins-det",
    reportNumber = "FERMILAB-PUB-24-0511-SQMS-V",
    doi = "10.1103/PhysRevD.110.022008",
    journal = "Phys. Rev. D",
    volume = "110",
    number = "2",
    pages = "022008",
    year = "2024"
}

@article{ORGAN:2024,
    author = "Quiskamp, Aaron P. and others",
    title = "{Near-quantum-limited axion dark matter search with the ORGAN experiment around 26{\,}{\,}{\ensuremath{\mu}}eV}",
    collaboration = "ORGAN",
    eprint = "2407.18586",
    archivePrefix = "arXiv",
    primaryClass = "hep-ex",
    doi = "10.1103/PhysRevD.111.095007",
    journal = "Phys. Rev. D",
    volume = "111",
    number = "9",
    pages = "095007",
    year = "2025"
}

@article{TASEH_DP_2025,
    author = "Chang, Yuan-Hann and Chiang, Cheng-Wei and Doan, Hien Thi and Houston, Nick and Li, Jinmian and Li, Tianjun and Wu, Lina and Zhang, Xin",
    title = "{Dark photon dark matter constraints and a tentative signal at the TASEH experiment}",
    eprint = "2507.00784",
    archivePrefix = "arXiv",
    primaryClass = "hep-ex",
    year = "2025",
    journal = ""
}

@article{Arias2012,
    author = "Arias, Paola and Cadamuro, Davide and Goodsell, Mark and Jaeckel, Joerg and Redondo, Javier and Ringwald, Andreas",
    title = "{WISPy Cold Dark Matter}",
    eprint = "1201.5902",
    archivePrefix = "arXiv",
    primaryClass = "hep-ph",
    reportNumber = "DESY-11-226, MPP-2011-140, CERN-PH-TH-2011-323, IPPP-11-80, DCPT-11-160",
    doi = "10.1088/1475-7516/2012/06/013",
    journal = "JCAP",
    volume = "06",
    pages = "013",
    year = "2012"
}

@book{Fabbrichesi2020,
    author = "Fabbrichesi, Marco and Gabrielli, Emidio and Lanfranchi, Gaia",
    title = "{The Dark Photon}",
    publisher = "Springer",
    year = "2020",
    doi = "10.1007/978-3-030-62519-1"
}
